\documentclass[prl,twocolumn,showpacs,aps]{revtex4}
\usepackage{graphicx}
\usepackage{bm}
\usepackage{amssymb}
\usepackage{amsmath}

\begin{document}

\title{Effect of superlattice modulation of electronic parameters on
superconducting density of states in cuprate superconductors}
\author{Kai-Yu Yang,$^{1}$ T. M. Rice,$^{1,2}$ and Fu-Chun Zhang$^{1}$}
\affiliation{$^{1}$ Centre of Theoretical and Computational Physics and Department of
Physics, The University of Hong Kong, Hong Kong\\
$^{2}$ Institut f\"{u}r Theoretische Physik, ETH Zurich, CH-8093 Z\"{u}rich,
Switzerland}
\date{Reveived 16 May 2007, revised manuscript received 11 August 2007}
\pacs{74.72.Hs, 74.20.-z, 74.62.Bf, 74.81.-g}

\begin{abstract}
Recent scanning tunneling microscopy on BSCCO 2212 has revealed a
substantial spatial supermodulation of the energy gap in the superconducting
state. We propose that this gap modulation is due to the superlattice
modulations of the atoms in the structure, and hence the parameters in a
microscopic model of the CuO$_{\text{2}}$ plane. The gap modulation is
estimated using renormalized mean field theory for a $t-t^{\prime }-J$ model
on a superlattice. The results compare well with experiment.
\end{abstract}

\maketitle

Scanning tunneling microscopy (STM) has become an important method to probe
the electronic structure in high-T$_{\text{c}}$ cuprates. \cite%
{Pan-Nature-01, Lang-Nature-02, Vershinin-Science-04, McElroy-science-05}
Recently STM has revealed a substantial modulation of the superconducting
energy gap\ along the a-axis in BSCCO 2212 samples. \cite{Davis modulation,
J. Slezak} The modulation appears in phase with the incommensurate periodic
superlattice modulation which originates in the Bi-O layers. \cite{BiO
structure, BiO structure2} Structure studies show that this lattice
modulation also affects the CuO$_{\text{2}}$ planes, and so will modify the
local electronic structure. Understanding this gap modulation may shed new
light on our understanding of the superconductivity mechanism of high-T$_{%
\text{c}}$ cuprates. In this paper, we analyze the effect of modulating the
parameters in a $t-t^{\prime }-J$ model on the superconducting state within
the renormalized mean field theory (RMF). \cite{Zhang-RMF} Our calculations
show that the observed superconducting gap modulation can be explained by a
reasonable choice for the parameter modulation within a $t-t^{\prime }-J$
model as a consequence of the short coherence length, although the precise
form of \textit{a priori} $t-t^{\prime }-J$ modulation will need more
concrete structure data. Zhu \cite{J X Zhu} has previously used a slave
boson approximation to examine the effect of defect and acceptor induced
parameter changes on superconductivity in the same model. Zhou \textit{et
al. }\cite{Sen Zhou} used RMF keeping $t$, $t^{\prime }$, and $J$ fixed
spatially but adding local Coulomb potentials to model dopant oxygen
acceptors close to the CuO$_{\text{2}}$\ planes to explain the nanoscale
inhomogeneity found in STM experiments. In this paper we restrict our model
to the superlattice structure. A somewhat different approach has been taken
by Andersen, Hirschfeld, and Slezak \cite{B. M. Andersen} who introduce a
periodic modulation of the pairing interaction in a conventional $d$-wave
BCS Hamiltonian to describe the superlattice.

Shortly after the discovery of the BSCCO superconductor, an incommensurate
lattice modulation with a period of $\approx $ 5 unit cells along its a axis
was found and characterized. \cite{BiO structure, BiO structure2} Note this
direction corresponds to a diagonal (1,1) direction in the CuO$_{\text{2}}$
square lattice. The determination of the detailed displacements and site
occupation that occur in this superlattice modulation is difficult to
undertake in a complex material such as BSCCO. While there is unanimity
among the various superlattice structures reported in the literature \cite%
{BiO structure, BiO structure2, D. Grebille, A. Yamamoto} that substantial
modulations occur in the key CuO$_{\text{2}}$ planes, there is no agreement
about their precise forms. \cite{no-precise-form} As a result, \textit{a
priori} estimates of the electronic parameter modulation within a $%
t-t^{\prime }-J$ model are still not possible at present.

We start from a $t-t^{\prime }-J$ model on a square lattice including the
no-double occupation constraint. Zhang and co-workers introduced the
Gutzwiller approximation, replacing the no-double occupation constraint by
the classical statistical weight factors $g_{t}^{i,j}$ and $g_{s}^{i,j}$ for
the hopping and spin-spin superexchange processes, respectively. \cite%
{Zhang-RMF} The RMF that they derived reproduces the results of Variational
Monte Carlo calculations on the full model quite well. Both have been proved
to be capable of explaining many experiments qualitatively and in some cases
even quantitatively \cite{RMF}. Decoupling the renormalized Hamiltonian
gives a mean field form in the presence of pairing%
\begin{eqnarray}
H &=&-\sum_{\left\langle i,j\right\rangle ,\sigma
}g_{t}^{i,j}t_{i,j}c_{i,\sigma }^{\dag }c_{j,\sigma }-\sum_{\left\langle
\left\langle i,j\right\rangle \right\rangle ,\sigma
}g_{t}^{i,j}t_{i,j}^{\prime }c_{i,\sigma }^{\dag }c_{j,\sigma }
\label{H_RMF} \\
&&-\mu \sum_{i,\sigma }c_{i,\sigma }^{\dag }c_{i,\sigma }-\frac{3}{8}%
\sum_{\left\langle i,j\right\rangle ,\sigma }g_{s}^{i,j}J_{i,j}[\chi
_{i,j}^{\ast }c_{i,\sigma }^{\dag }c_{j,\sigma }+\mathbf{H}.c.]  \notag \\
&&-\frac{3}{8}\sum_{\left\langle i,j\right\rangle ,\sigma
}g_{s}^{i,j}J_{i,j}[\Delta _{i,j}^{\ast }\left( c_{i,\uparrow
}c_{j,\downarrow }-c_{i,\downarrow }c_{j,\uparrow }\right) +\mathbf{H}.c.] 
\notag
\end{eqnarray}%
where the operator $c_{i,\sigma }^{\dag }\left( c_{i,\sigma }\right) $
creates (annihilates) an electron with spin $\sigma $ on the $i$ th lattice
site.$\ t_{i,j}$ and $t_{i,j}^{\prime }$ are the hopping integrals for
nearest neighboring (NN) sites $\left\langle i,j\right\rangle $ and
next-nearest neighboring (NNN) sites $\left\langle \left\langle
i,j\right\rangle \right\rangle $, respectively. $J_{i,j}$ is the NN
spin-spin coupling constant, $\mu $ is the chemical potential, and $\chi
_{i,j}=\sum_{\sigma }\left\langle c_{i,\sigma }^{\dag }c_{j,\sigma
}\right\rangle $, $\Delta _{i,j}=\left\langle c_{i,\uparrow }c_{j,\downarrow
}-c_{i,\downarrow }c_{j,\uparrow }\right\rangle $ are the local NN
particle-hole and particle-particle pairing fields. The Gutzwiller
renormalization factors $g_{t}^{i,j}$ and $g_{s}^{i,j}$ depend on the local
doping as, \cite{Q.H. Wang} 
\begin{equation}
g_{t}^{i,j}=\sqrt{\frac{2\delta _{i}}{1+\delta _{i}}\frac{2\delta _{j}}{%
1+\delta _{j}}},\text{ \ \ \ }g_{s}^{i,j}=\frac{4}{\left( 1+\delta
_{i}\right) \left( 1+\delta _{j}\right) },  \label{g-factor}
\end{equation}%
where $\delta _{i}=1-n_{i}$ is the on-site hole concentration with $%
n_{i}=\sum_{\sigma }\left\langle c_{i,\sigma }^{\dag }c_{i,\sigma
}\right\rangle $.

\begin{figure}[tbp]
\includegraphics [width=8.5cm,height=4.475cm] {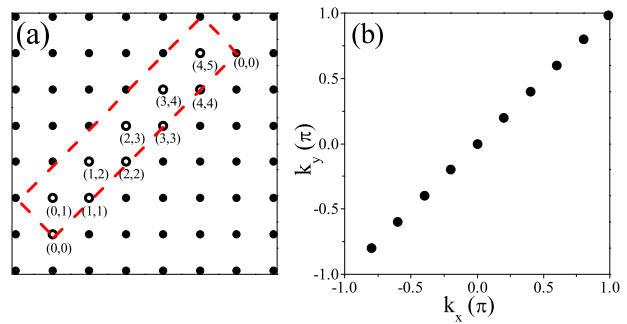}
\caption{(Color online) (a) The supercell structure. The black open dots
present the sites in a unit cell with the coordinate inside the cell
indicated. (b) The reciprocal lattice for a unit cell.}
\label{lattice}
\end{figure}

As discussed above, the incommensurate lattice modulation has a period of $%
\approx $ 5 unit cells along the diagonal direction. In this paper we
examine an electronic superlattice structure with a commensurate
periodicity, i.e., each unit cell consists of 10 sites as illustrated in
Fig. \ref{lattice}(a). This superlattice has a period of $5\sqrt{2}$ lattice
constants along the (1,1) direction and $\sqrt{2}$ perpendicular to it. The
reciprocal lattice is shown in Fig. \ref{lattice}(b). We use $\mathbf{I}%
\left( \mathbf{J}\right) $ to index the supercell and $i\left( j\right) $
for the original lattice. The lattice constant is set as the length unit with%
$\ \mathbf{\tau }=\left( 0,\pm 1\right) $ or $\left( \pm 1,0\right) $ for
NN, and $\mathbf{\Gamma }=\left( \pm 1,\pm 1\right) $ for NNN. Here we
assume a simple cosine modulation, consistent with the dominant lattice
modulation observed in Bi-O, Sr-, and CuO$_{\text{2}}$ layers, 
\begin{eqnarray}
t_{\mathbf{I},\mathbf{I}+\mathbf{\tau }}/t_{0} &=&1+\frac{A}{2}\left[ \cos {(%
\mathbf{Q}\cdot \mathbf{R}_{\mathbf{I}}}+\cos {(\mathbf{Q}\cdot \mathbf{R}_{%
\mathbf{I}+\mathbf{\tau }})}\right] ,  \label{modulation} \\
J_{\mathbf{I},\mathbf{I}+\mathbf{\tau }}/J_{0} &=&\left( t_{\mathbf{I},%
\mathbf{I}+\mathbf{\tau }}/t_{0}\right) ^{2},  \notag \\
t_{\mathbf{I},\mathbf{I}+\mathbf{\Gamma }}^{\prime }/t_{0}^{\prime } &=&1+%
\frac{B}{2}\left[ \cos {(\mathbf{Q}\cdot \mathbf{R}_{\mathbf{I}+\Gamma _{x}})%
}+\cos {(\mathbf{Q}\cdot \mathbf{R}_{\mathbf{I}+\Gamma _{y}})}\right] , 
\notag
\end{eqnarray}%
where $\mathbf{Q}=\left( \pi /5,\pi /5\right) $, and $\mathbf{R}_{\mathbf{I}}
$ is the lattice coordinate within a unit cell. The average parameters are
set to be $t_{0}=300$ meV, $J_{0}/t_{0}=0.3$, $t_{0}^{\prime }/t_{0}=-0.3$,
comparable to the experimentally observed values. A quadratic relation
between $t$ and $J$ is preserved unless specified otherwise. The Hamiltonian
in Eq. (\ref{H_RMF}) with the supercell structure is solved
self-consistently. The local density of states (LDOS) $N_{\mathbf{I}}\left(
\omega \right) $ is determined by a Fourier transformation of the local
Green's function $iG_{\mathbf{I},\sigma }=\left\langle T_{\tau }[c_{\mathbf{I%
},\sigma }^{\dag }\left( \tau \right) c_{\mathbf{I},\sigma }\left( 0\right)
]\right\rangle $ multiplied by a renormalization factor $g_{t}^{\mathbf{I},%
\mathbf{I}}$, where $T_{\tau }$ is the time ordering operator.

Our simple model with modulations of $t$, $t^{\prime }$, and $J$ can
reproduce many of the most prominent features of the STM experiments. The
variation of the gap (about 10\%) within a wide doping range is reproduced
by a reasonable choice of the electronic parameter modulation in our theory.
It is worth noting that our model gives a very weak doping dependence of the
relative gap variance, which agrees well with the data recently reported by
the Cornell group. \cite{Davis modulation, J. Slezak} In addition, the
low-energy subgap spectra are spatially homogeneous despite the
inhomogeneity in the gap. Further the coherent peaks are always almost
symmetrically located around zero bias. The gap is negatively correlated
with the local doping concentration. In our calculation, we have normalized
the gap value obtained from the RMF by a factor of 2 to bring them in line
with the more accurate Variational Monte Carlo results. \cite{RMF}

To compare with the recent STM\ experiments, three typical doping levels
with $\delta $=0.13 (UD), 0.16 (OP), 0.19 (OD) are considered. In our
numerical calculation, a 1000 $\times $ 400 supercell structure is
introduced. The LDOS presented here is filtered by fast Fourier
transformation with an energy window of 11 meV. We have studied the effects
of various possible modulations of parameters $t$, $t^{\prime }$ and $J$ to
the LDOS and to the local gap values. The effects are qualitatively similar,
although it is much less sensitive to vary $t^{\prime }$ than $t$ or $J$.

\begin{figure}[tbp]
\includegraphics [width=8.5cm,height=9.5cm] {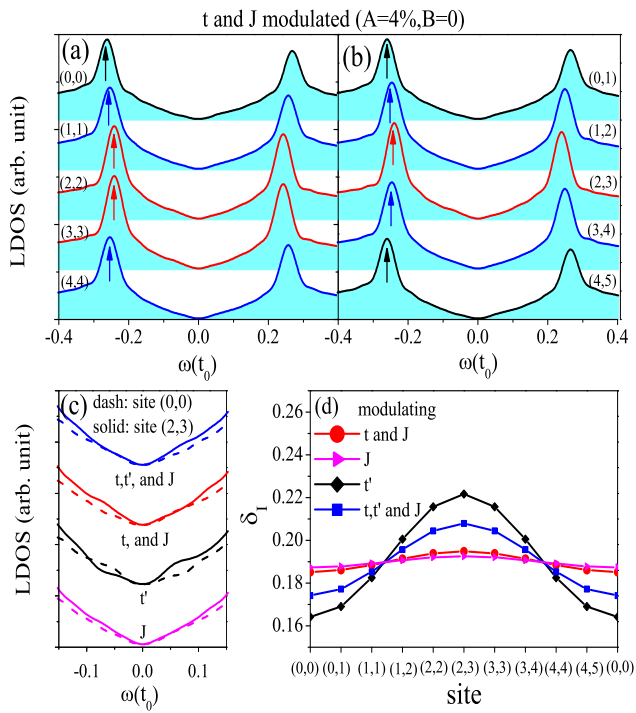}
\caption{(Color online) Panels (a)-(b) On-site LDOS with $t$ and $J$
modulated [$A=4\%$,B=0 in Eq. (\protect\ref{modulation})] with the peak
positions indicated by arrows. The energy $\protect\omega $ is in unit of $%
t_{0}$ ($=300$ meV). Panel (c) The LDOS at low energies at two supercell
sites (0,0) and (2,3), and panel (d) for on-site local hole concentration.
The doping is $\protect\delta =0.19$. In (c)-(d), $A=4\%$ in the modulations
of $t$ and $J$ modulation (red), and $J$ only with $t$ fixed (magenta), $%
B=45\%$ in the modulation of $t^{\prime }$ (black), and $A=2\%$ and $B=22.5\%
$ in the modulation of $t$, $t^{\prime }$, and $J$ (blue). }
\label{LDOS}
\end{figure}

\begin{figure}[tbp]
\includegraphics [width=8.5cm,height=9.5cm] {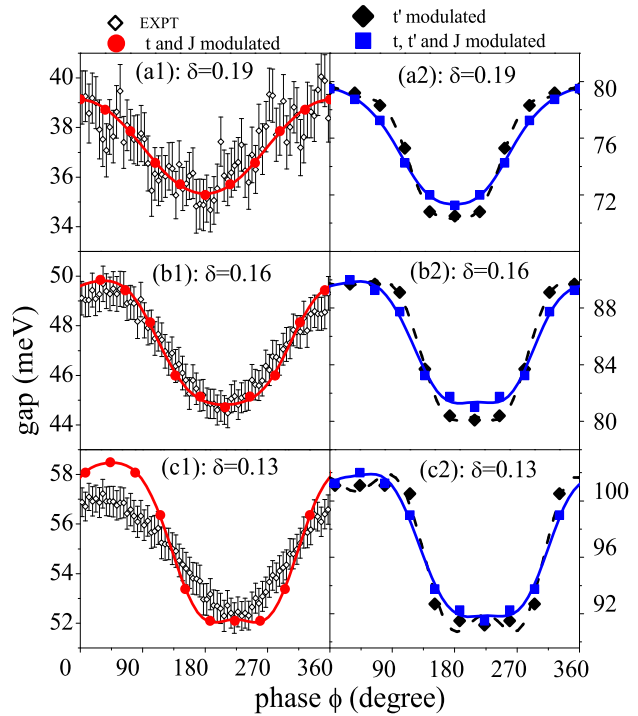}
\caption{(Color online) The gap modulation due to electronic parameter
modulations. The phase $\protect\phi $ characterizes the electronic
structure supermodulation which can be described by a cosine form.
Experimentally the supermodulation phase mapping $\protect\phi \left(
r\right) $ is consistent with the original topograph $Z(r)$. (Ref. 
\protect\cite{Davis modulation, J. Slezak}) The theoretical gap modulation
vs $\mathbf{Q\cdot R}_{I}$ (the range is linearly scaled to 360$^{\circ }$)
is shifted along the horizonal direction to match the gap minimum and
maximum loci. The open diamonds with error bars in panels (a1)-(c1)
represent STM experimental data (Ref. \ \protect\cite{Davis modulation, J.
Slezak}) for three samples (OD,OP,UD) with the energy gap $\Delta =37$, $47$%
, $55$ meV, respectively. Solid symbols are from our calculations, red
circles for $t$ and $J$ modulation ($A=4\%$, $B=0$), black diamonds for $%
t^{\prime }$ modulation ($A=0$, $B=45\%$), and blue squares for $t$, $%
t^{\prime }$ and $J$ modulations ($A=2\%$, $B=22.5\%$). Color curves show
analytic interpolations of our calculations between lattice sites. }
\label{gap}
\end{figure}

First we consider modulation of $t$ and $J$ while keeping $t^{\prime }$
unchanged, where the modulation amplitudes $A=4\%$ and $B=0$. The resulting
variances of $t$ and $J$ are $\sim $7 and $\sim $14\%, respectively. The
results shown in Fig. \ref{LDOS} are for doping 0.19, which are
representative for all the three hole concentrations studied here. The LDOS
is shown in Figs. \ref{LDOS}(a) and \ref{LDOS}(b). At low energy there is a
clear homogeneous \textquotedblleft V\textquotedblright\ shape [panel(c)]
indicating the nodal structure ($v_{F}$, $v_{\Delta }$) is robust against
this electronic modulation. The homogeneity in LDOS at low energy was also
found by Wang \textit{et al.,} \cite{Q.H. Wang} in studies of disorder
effects. The coherent peaks with lower height are located at higher energy
and the spectral weight suppressed at low energy is transferred to high
energy near the band edge. Note that there is a multipeak character of the
LDOS which is smeared by fast Fourier transformation. Similar multipeak
character has also been observed in a recent STM experiment with high-energy
resolution up to 2meV. \cite{A. C. Fang} The modulation of the
superconducting energy gap [shown in Fig. \ref{gap}(a1)] has a variation
about 10\% for this parameter choice, which matches well with recent STM
experiments. The doping variance achieved here is about 5\%, i.e., 0.185$-$%
0.195 [shown in Fig. \ref{LDOS}(d)], comparable with the value of 7\%
recently reported in the STM experiments. \cite{McElroy-science-05,
Nunner-prl-05} The negative correlation between the gap and the doping
concentration is reproduced. In addition, there is a small deviation from
the original $d$-wave symmetry due to the symmetry breaking imposed by the
supercell structure orientation rotated by 45$^{\circ }$. In Figs. \ref{gap}%
(a1)-\ref{gap}(c1) the modulations of the gap for different dopings are
shown compared with the STM results. The phase $\phi $ serving as the
horizonal axis characterizes the electronic structure supermodulation which
can be described by a consine form. The supermodulation phase mapping $\phi
\left( r\right) $ is consistent with the observed original topograph $Z(r)$. 
\cite{Davis modulation, J. Slezak} The theoretical gap modulation vs $%
\mathbf{Q\cdot R}_{I}$ (the range is linearly scaled to 360$^{\circ }$) is
shifted along the horizonal direction to match the gap minimum and maximum
loci. Keeping the $t$ and $J$ modulation constant, a very weak doping
dependence of the variance of the gap modulation is observed. Both the form
and the magnitude of the gap modulation agree well with the experimental
results on samples ranging between the underdoped and overdoped regime. The
decreased agreement happened at low doping ($\delta =0.13$) is mainly due to
the larger peak-to-peak distance of the on-site LDOS which possesses a
multipeak structure before fast Fourier transformation.

We have also considered the superlattice modulations of $t^{\prime }$ only, $%
J$ only, and a combined modulation of $t-t^{\prime }-J$. The low-energy LDOS
and the local hole density for the various choices of the modulated
parameters are shown in Figs. \ref{LDOS}(c) and \ref{LDOS}(d), respectively.
In Figs. \ref{gap}(a)-\ref{LDOS}(c), we show the gap modulation obtained
from the various choices of parameter modulations. It requires a substantial
modulation of $t^{\prime }$, i.e., $A=0$ and $B=45\%$, corresponding to a
range for $t^{\prime }=0.165-0.435$, to have similar gap modulation with $%
\sim $10\% change in gap. The large variation in local hole density is due
to the large modulation of $t^{\prime }$ required to obtain this modulation
of the gap. By modulating $J$ ($A=4\%$) alone with the quadratic relation
between $t$ and $J$ released, we have found only a slight difference from
the case with both $t$ and $J$ modulated simultaneously. Hence the results
for modulating $J$ alone are omitted in Fig. \ref{gap} due to the close
similarity to modulating $t$ and $J$ simultaneously. The low-energy behavior
becomes more homogeneous and the variation of the doping concentration
shrinks as shown in Figs. \ref{LDOS}(c) and \ref{LDOS}(d). Zhu who examined
the nanoscale inhomogeneity \cite{J X Zhu} presented a similar picture
arguing that the presence of the randomly distributed out-of-plane dopant
oxygens will lead to a change in the superexchange strength. From the cases
studied above, we can reach a conclusion that, with similar variance of the
superconducting energy gap, a modulation of $t^{\prime }$ leads to a larger
modulation of doping concentration while $J$ and/or $t$ tend to give a
better low-energy homogeneity. We have examined a combined modulation for a
choice of $A=2\%$ and $B=22.5\%$. The LDOS, superconducting energy gap,
doping concentration, all show an additive behavior from the individual
modulation of $A$ and $B$ with little interference, and the low-energy LDOS
becomes more homogeneous. From our current study it is obvious that
modulation of the electronic parameters $t$, $t^{\prime }$, and $J$ is able
to generate the superlattice modulation phenomenon recently observed in STM. 
\cite{Davis modulation, J. Slezak}

Pavarini \textit{et al. }\cite{O K Andersen} have proposed that the
variation of the transition temperature $T_{c}$ between different cuprate
families is controlled by the Cu-O3 apical distance through its effect on
the\ NNN hopping parameter $t^{\prime }$. The energy gap estimated by the
RMF theory shows a similar dependence on $t^{\prime }$, namely, the gap
increases with increasing values of $\left\vert t^{\prime }\right\vert $ at
constant hole doping in agreement with the increase in $T_{c}$ argued by
Pavarini \textit{et al}. \cite{O K Andersen} It is therefore appealing to
argue that this is the controlling feature of the superlattice modulation.
However, there is a real problem with this interpretation. Slezak in his
thesis \cite{J. Slezak} finds that if he uses the Cu-O3 distance modulation
reported by Yamamoto \textit{et al. }\cite{A. Yamamoto} there is a negative
correlation, i.e. smaller values of the gap occur at larger Cu-O3 distances.
This is opposite to the conclusion of Pavarini \textit{et al.} \cite{O K
Andersen} that $\left\vert t^{\prime }\right\vert $ scales with the Cu-O3
distance. Thus the simplest interpretation that modulation of $\left\vert
t^{\prime }\right\vert $ is the controlling factor cannot be justified.
Instead one cannot directly relate the reported structural data and the gap
modulation. One can say, however, that the size of the modulations in the
electronic parameters needed to explain the energy gap modulations are quite
reasonable even if we cannot identify the specific values of individual
parameter modulations.

Lastly we comment on a couple of related issues. Slezak in his thesis \cite%
{J. Slezak} also reports on modulations of the bosonic mode $\Omega $
accompanying the superlattice modulations. However, as he points outs the
isotope shift of $\Omega $ has no effect on the gap which leads to the
conclusion that both vary as a consequence of the structural modulation as
was also suggested by Pilgram \textit{et al}. \cite{S. Pilgram} A second
feature of the BSCCO cuprates is the substantial disorder in the STM gap
maps. \cite{McElroy-science-05} This is characterized by regions with
enhanced gap values associated with a high energy resonance in the STM
spectra which has been interpreted as originating in a nearby dopant O$^{%
\text{2-}}$ ion. \cite{McElroy-science-05} This leads to a problem since an
enhanced hole density would be expected around a O$^{\text{2-}}$ ion but the
enhanced gap suggests a locally reduced hole density. Several proposals have
been made to reconcile theses two features \cite{J X Zhu, Sen Zhou,
Nunner-prl-05}. However, Yamamoto \textit{et al.} \cite{A. Yamamoto} and
also Eisaki \textit{et al.} \cite{H. Eisaki} conclude that in BSCCO a
substantial number of Bi ions substitute for the Sr ions on planes adjacent
to the CuO$_{\text{2}}$ planes. This suggests to us as it did to Eisaki 
\textit{et al.} \cite{H. Eisaki}, that since Bi$^{\text{3+}}$ acts as a
donor ion, the regions of enhanced gap should be associated with Bi$^{\text{%
3+}}$ donors rather than O$^{\text{2-}}$ acceptors.

We are very grateful to S\'{e}amus Davis and to S. Uchida for stimulating
discussion. This work was supported by the RGC grant and Centre of
Theoretical and Computational Physics of HKSAR, RGC Central Allocation Grant
of HKSAR, Visiting Professorship at The University of Hong Kong, and the
MANEP program of the Swiss National Foundation.

\end{document}